\documentclass[aps,prb,preprint,showpacs,groupedaddress]{revtex4}
\usepackage{amsmath}
\usepackage{amssymb}
\usepackage{graphicx}
\usepackage{dcolumn}
\usepackage{bm}
\usepackage{subfigure}

\begin{document}

\title{The elastic constants of solid $^{4}$He under pressure: \\
                  a diffusion Monte Carlo study}
\author{C. Cazorla$^{\rm a}$, Y. Lutsyshyn$^{\rm b}$ and J. Boronat$^{\rm c}$}
\affiliation{
$^{\rm a}$ Institut de Ci$\grave{e}$ncia de Materials de Barcelona (ICMAB-CSIC), 
08193 Bellaterra, Spain \\
$^{\rm b}$ Institut f\"{u}r Physik, Universit\"{a}t Rostock, 18051 Rostock, Germany \\ 
$^{\rm c}$ Departament de F\'{i}sica i Enginyeria Nuclear, Universitat Polit\`{e}cnica 
de Catalunya, Campus Nord B4-B5, E-08034, Barcelona, Spain 
} 

\begin{abstract}

We study the elasticity of perfect $^{4}$He at zero-temperature
using the diffusion Monte Carlo method and a realistic semi-empirical 
pairwise potential to describe the He-He interactions.
In particular, we calculate the value of the elastic 
constants of hcp helium $\lbrace C_{ij} \rbrace$ as a function of 
pressure up to $\sim 110$~bar. 
It is found that the pressure dependence of all five non-zero 
$\lbrace C_{ij} \rbrace$ is linear and we provide an accurate 
parametrization of each of them. 
Our elastic constants results are compared to previous variational
calculations and low-temperature measurements and in general  
we find notably good agreement among them.
Furthermore, we report $T = 0$ results for the Gr\"{u}neisen parameters, 
sound velocities and Debye temperature of hcp $^{4}$He. 
This work represents the first of a series of computational studies
aimed at thoroughly characterizing the response of solid helium 
to external stress-strain. 


\end{abstract}

\maketitle

\section{Introduction}
\label{sec:intro}

The behavior of solid $^{4}$He, and of quantum crystals 
in general (e.g., H$_{2}$ and Ne), is exceptionally 
so rich that despite having been investigated 
for more than about eight decades is to this day not yet 
completely understood. One example of helium's intriguing
nature is its elasticity. Experimental studies on the elastic 
properties of hcp $^{4}$He were already conducted by Wanner, 
Crepeau and Greywall in the early 
seventies.~\cite{wanner70,crepeau71,greywall71}
Those original works consisted of series of sound-velocity 
measurements performed at thermodynamic conditions 
relatively close to the stability domain of the 
liquid, namely $T \sim 1$~K and $25 \le P \le 50$~bar. 
With the advance of time and technology cryogenic 
and crystal growth techniques have been improved so notably 
that at present is possible to analyze practically defect-free 
$^{4}$He samples at just few mK in the laboratory.           
Recently, Beamish and collaborators have developed a new 
experimental technique that has allowed them to measure 
directly the shear modulus $\mu$ of hcp $^{4}$He under 
extremely low strains and frequencies.~\cite{day07,syshchenko09}
The temperature dependence of $\mu$ within the temperature interval 
$0.5 \le T \le 0.01$~K has been determined and a striking 
resemblance with non-classical rotational inertia 
(NCRI) data obtained in torsional oscillator 
experiments~\cite{kim04a,kim04b} has been unravelled. 
Specifically, the value of the NCRI and shear modulus of  
helium increases about $2$~\% as the temperature is lowered 
down to $0.01$~K. Despite Beamish \emph{et al.}'s 
findings have been initially rationalized in terms of 
pinning~(unpinning) of dislocations induced by the presence of 
static~(mobile) $^{3}$He impurities,~\cite{day07,day10,day09,rojas10} 
it remains to be clarified whether the cited experimental 
similarities must be regarded simply as coincidental or contrarily 
correspond to manifestations of a same and unique quantum phenomenon 
known as supersolidity.~\cite{kim10,kim11,iwasa10}         

Simulation techniques have been demonstrated as invaluable tools 
for predicting and accurately characterizing the energetic and 
structural properties of quantum 
solids.~\cite{cazorla08a,cazorla08b,boronat04,cazorla04,cazorla08c,cazorla10} 
Nevertheless, computational studies on the elasticity of 
solid $^{4}$He so far have been very infrequent. 
To the best of our knowledgement, 
there exists only one recent work by Pessoa \emph{et al.} in which 
the shear modulus of solid helium has been explicitly calculated
from first-principles.~\cite{pessoa10}
This computational scarcity strongly contrasts with research 
done in other fields like classical solid state theory and/or
high pressure physics, where estimation 
of the elastic properties of materials (e.g. strain-stress tensor,
Gr\"{u}neisen parameters, vibrational phonon frequencies, etc.) 
is a standard.~\cite{cazorla09,stixrude99,cazorla08d,gillan06} 
The likely explanation for such a constrast (besides 
no particular enthusiasm in the matter prior to Beamish \emph{et al.} 
experiments) are the technical difficulties encountered in 
modeling of bosonic quantum effects, namely  
atomic delocalization, anharmonicity and particle exchanges. 
These quantum atomic effects are indeed crucial to comprehend the 
nature of solid $^{4}$He at low temperatures and, as a matter 
of fact, customary harmonic and quasi-harmonic computational 
approaches~\cite{baroni01,alfe09,kresse95} 
can not be used to obtain a reliable picture of it. 

Here, we present a computational study of the elastic properties
of perfect (e.g. free of defects) solid $^{4}$He in the hcp 
structure based on the diffusion Monte Carlo approach. 
This work is intended to improve our understanding of how  
solid helium reacts to external strains/stresses, and further extends 
the work initiated by Pessoa \emph{et al.}~\cite{pessoa10} 
In particular, we provide the zero-temperature dependence of helium elastic 
constants and related quantities (e.g. sound velocities, 
Gr\"{u}neisen parameters and the Debye temperature) 
on pressure up to $\sim 110$~bar. Our results are compared to experimental 
data and other calculations whenever is possible and, as it will be shown 
later on, good agreement among them is generally found. The computational method that
we employ is fully quantum and virtually exact (i.e., in principle 
only affected by statistical uncertainties) so that from a technical point 
of view our study also represents an improvement with respect to previous 
first-principles work~\cite{pessoa10} based on variational Monte Carlo 
calculations (i. e., subjected to statistical and importance sampling biases).        

The remainder of the article is as follows. In Section~\ref{sec:theory}, 
we review the basics of elasticity in hcp crystals and provide 
the details of our computational method and strategy.    
In the following section, we present our results along with some discussions.
Finally, we summarize the main conclusions obtained and comment on prospective 
work in Section~\ref{sec:summary}.

\section{Theory and computational details}
\label{sec:theory}

\begin{figure}
\centerline{
\includegraphics[width=0.8\linewidth]{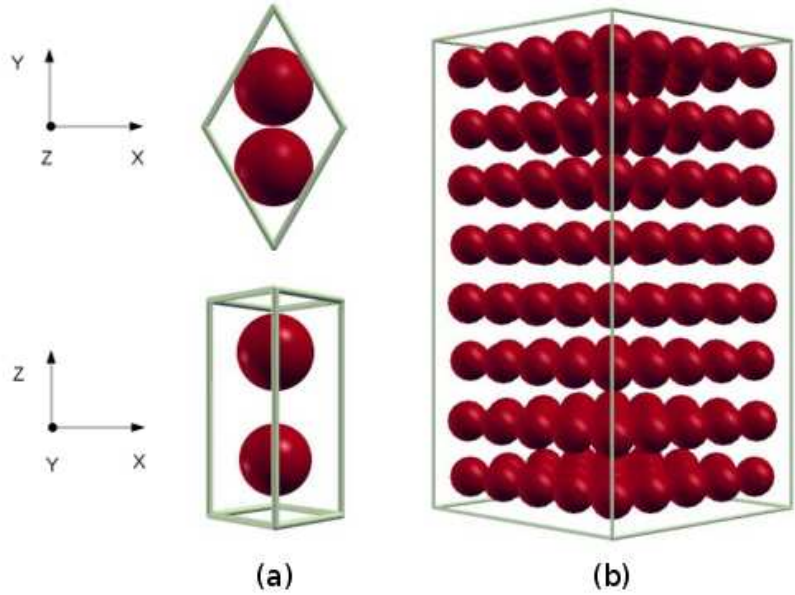}}%
\caption{(a)~Representation of the hcp unit cell with
primitive translational vectors ${\bf a_{1}}$, 
${\bf a_{2}}$ and ${\bf a_{3}}$, and 
two-atoms basis set (see text). (b)~Sketch of the $200$-atoms 
supercell used in the pure shear calculations, built by 
replicating the hcp unit cell $5 \times 5 \times 4$ times 
along the primitive translational vectors ${\bf a_{1}}$, 
${\bf a_{2}}$ and ${\bf a_{3}}$, respectively.}
\label{fig:supercell}
\end{figure}

\subsection{Elastic constants}
\label{subsec:elasticcte}

For small strains, the zero-temperature energy of a crystal can be 
expressed as
\begin{equation}
E = E_{0} + \frac{1}{2} V_{0} \sum^{6}_{i,j = 1} C_{ij} s_{i} s_{j}~, 
\label{eq:energy}
\end{equation}  
where $V_{0}$ and $E_{0}$ are the volume and internal energy
of the undistorted solid, 
$\lbrace C_{ij} \rbrace$ the elastic constants 
and $\lbrace s_{i} \rbrace$ the strain components defined such 
that $s_{1}$, $s_{2}$ and $s_{3}$ are fractional increases in the $x$, 
$y$ and $z$ directed axes, and $s_{4}$, $s_{5}$ and $s_{6}$ angular 
increases of the $xy$, $xz$ and $yz$ angles.~\cite{barron65,wallace72,king70} 
The symmetry properties of the crystal under consideration define
the number of elastic constants which are non-zero. 
For hcp crystals, only five elastic contants are different from zero,  
namely  $C_{11}$, $C_{12}$, $C_{33}$, $C_{13}$ and $C_{44}$,    
where $C_{44}$ is commonly known as the shear modulus and abbreviated 
$\mu$.     
In order to calculate these five non-zero elastic constants is necessary
to compute the second derivative of the internal energy of the crystal 
with respect to the strain tensor $\sigma_{ij}$. For this, the hcp 
crystal must be first considered in full symmetry, that is, expressed in 
terms of its unit cell with primitive translational vectors 
${\bf a_{1}} = a \left(+\frac{1}{2} {\bf i} + \frac{\sqrt{3}}{2} {\bf j} \right)$,
${\bf a_{2}} = a \left(-\frac{1}{2} {\bf i} + \frac{\sqrt{3}}{2} {\bf j} \right)$
and ${\bf a_{3}} = c {\bf k}$ (where $a$ and $c$ are the lattice parameters 
in the basal plane and along the $z$ axis respectively, 
and ${\bf i}$, ${\bf j}$ and ${\bf k}$ correspond to the 
usual unitary Cartesian vectors), and two-atom basis set 
${\bf r_{1}} = \frac{1}{2}{\bf a_{1}} + \frac{1}{3}{\bf a_{2}} + \frac{2}{3}{\bf a_{3}}$ 
and ${\bf r_{2}} = ( 0 , 0 , 0)$ (see Fig.~\ref{fig:supercell}).

The relationships between the hcp elastic constants $\lbrace C_{ij} \rbrace$
and applied strain were determined long-time ago within the framework
of elasticity theory, those being~\cite{barron65,wallace72,king70} 
\begin{equation}
K = -V \left(\frac{\partial P}{\partial V}\right)_{V = V_{0}} = \frac{C_{33}\left( C_{11} + C_{12}\right) - 2C_{13}^{2}}{C_{11} + C_{12} + 2C_{33} - 4C_{13}}~, 
\label{eq:bulk}
\end{equation}

\begin{equation}
-V \left(\frac{\partial \ln{c/a}}{\partial V}\right)_{V = V_{0}} = \frac{C_{33} - C_{11} - C_{12} + C_{13}}{C_{11} + C_{12} + 2C_{33} - 4C_{13}}~,
\label{eq:lnca}
\end{equation}

\begin{equation}
C_{0} = C_{11} + C_{12} + 2C_{33} - 4C_{13}~,
\label{eq:c0}
\end{equation}

\begin{equation}
C_{66} = \frac{1}{2}\left( C_{11} - C_{12} \right)   
\label{eq:c1}
\end{equation}
and
\begin{equation}
C_{44} = C_{44}~.   
\label{eq:c44}
\end{equation}

Equations~(\ref{eq:bulk}) and~(\ref{eq:lnca}) correspond to
homogeneous strains that change the volume and 
shape of the hcp unit cell. The dependence 
of pressure $P$ and $c/a$ ratio on volume can 
be readily obtained from standard equation of 
state calculations. 
On the other hand, equations~(\ref{eq:c0}), (\ref{eq:c1}) 
and~(\ref{eq:c44}) represent heterogeneous strains that keep the 
volume of the hcp unit cell unaltered. In order to calculate 
the value of the pure shears $C_{0}$, $C_{66}$ and $C_{44}$ is 
necessary to compute the variation of the internal energy 
of the equilibrium structures with respect to certain 
deformations, which can be written as transformed primitive 
translational vectors. In the $C_{0}$ case, these are
\begin{eqnarray}
{\bf a_{1}^{0}} = a\epsilon^{-1} \left(+\frac{1}{2} {\bf i} + \frac{\sqrt{3}}{2} {\bf j} \right) \nonumber \\
{\bf a_{2}^{0}} = a\epsilon^{-1} \left(-\frac{1}{2} {\bf i} + \frac{\sqrt{3}}{2} {\bf j} \right) \nonumber \\
{\bf a_{3}^{0}} = c\epsilon^{2} {\bf k}~,
\label{eqn:c0vec}
\end{eqnarray}  
where $\epsilon = (1 + \eta)^{1/2}$,  
$C_{0} = \frac{2}{V_{0}}\left( \frac{\partial^{2} E}{\partial \eta^{2}} \right)_{V=V_{0}}$
and the equilibrium condition is satisfied at $\eta = 0$.
For $C_{66}$, we have 
\begin{eqnarray}
{\bf a_{1}^{66}} = a\gamma^{1/2} \left(+\frac{1}{2} {\bf i} + \gamma^{-1} \frac{\sqrt{3}}{2} {\bf j} \right) \nonumber \\
{\bf a_{2}^{66}} = a\gamma^{1/2} \left(-\frac{1}{2} {\bf i} + \gamma^{-1} \frac{\sqrt{3}}{2} {\bf j} \right) \nonumber \\
{\bf a_{3}^{66}} = c{\bf k}~,
\label{eqn:c1vec}
\end{eqnarray}
where 
$C_{66} = \frac{1}{V_{0}}\left( \frac{\partial^{2} E}{\partial \gamma^{2}} \right)_{V=V_{0}}$
and the equilibrium condition is satisfied at $\gamma = 1$.
And finally for $C_{44}$,
\begin{eqnarray}
{\bf a_{1}^{44}} = a \left(+\frac{1}{2} {\bf i} + \frac{\sqrt{3}}{2} {\bf j} - \frac{\phi}{2} {\bf k} \right) \nonumber \\
{\bf a_{2}^{44}} = a \left(-\frac{1}{2} {\bf i} + \frac{\sqrt{3}}{2} {\bf j} - \frac{\phi}{2} {\bf k} \right) \nonumber \\
{\bf a_{3}^{44}} = c{\bf k}~,
\label{eqn:c44vec}
\end{eqnarray}
where 
$C_{44} = \frac{1}{V_{0}}\left( \frac{\partial^{2} E}{\partial \phi^{2}} \right)_{V=V_{0}}$
and the equilibrium condition is satisfied at $\phi = 0$.

Once the value of the bulk modulus $K$ and quantities 
$\partial \ln{(c/a)} / \partial V$, 
$C_{0}$, $C_{66}$ and $C_{44}$ is determined, one can 
calculate the five corresponding $C_{ij} \neq 0$ 
hcp elastic constants straightforwardly by solving 
the non-linear system of equations defined by
Eqs.~(\ref{eq:bulk})-(\ref{eq:c44}).

\subsection{Diffusion Monte Carlo}
\label{subsec:dmc}

The fundamentals of the diffusion Monte Carlo (DMC) method have been
reviewed with detail in other works~\cite{hammond94,guardiola98,ceperley86} 
so for brevity's sake we recall here only the essential ideas. 

In the DMC approach, the time-dependent Schr\"odinger equation 
of a quantum system of $N$ interacting particles is solved 
stochastically by simulating the time evolution of the Green's function 
propagator $e^{-\frac{i}{\hbar} \hat{H} t}$ in imaginary time 
$\tau$. For $\tau \to \infty$, sets of configurations (walkers) 
$\lbrace {\bf R}_i \equiv {\bf r}_1,\ldots,{\bf r}_N \rbrace$ 
rendering the probability distribution function 
($\Psi_0 \Psi$) are generated, where $\Psi_0$ is the true ground-state
wave function of the system and $\Psi$ the trial wave function used for 
importance sampling. Within DMC, virtually exact results (i.e., 
subjected to statistical uncertainties only) are obtained 
for the ground-state total energy and related quantities
in bosonic quantum systems.~\cite{aclaration,barnett91,casulleras95} 
It is worth noticing that despite asymptotic DMC values
do not depend on the choice of the trial wave function, 
the corresponding algorithmic efficiency is tremendously 
affected by the quality of $\Psi$.

We are interested in studying the ground-state of perfect hcp $^4$He, which 
we assume to be governed by the Hamiltonian  
$H= -\frac{\hbar^2}{2m_{\rm He}} \sum_{i=1}^{N} \nabla_i^2 + \sum_{i<j}^{N} V_{\rm He-He}(r_{ij})$
where $m_{\rm He}$ is the mass of an helium atom, 
$N$ the number of particles and $V_{\rm He-He}$ 
the semi-empirical pairwise potential due to Aziz \emph{et al}.~\cite{aziz2} 
It is worth noting that this two-body potential provides an excellent
description of the He-He interactions, including
weak long-ranged van der Waals forces, over all the pressure 
range considered in this work.~\cite{boro94,cazorla08a} 

The trial wave function that we use for importance 
sampling $\Psi_{\rm SNJ}$ simultaneously reproduces crystal ordering 
and Bose-Einstein symmetry (that is, remains unchanged under the 
permutation of atoms). This model wave function was recently 
introduced in Ref.~[\onlinecite{cazorla09b}] and it reads 
\begin{equation}
\Psi_{\rm SNJ}({\bf r}_1,\ldots,{\bf r}_N) = \prod_{i<j}^{N} f(r_{ij}) 
\prod_{J=1}^{N} \left( \sum_{i=1}^{N} g(r_{iJ}) \right)~,
\label{snjtrial}
\end{equation}
where the index in the second productory runs over perfect lattice
position vectors (sites).
In previous works, we have demonstrated that $\Psi_{\rm SNJ}$ provides 
an excellent description of the ground-state properties of bulk hcp 
$^{4}$He~\cite{cazorla09b} and quantum solid 
films.~\cite{cazorla08b,cazorla10,gordillo11}
The key ingredient for this progress stays in the  
$\Psi_{\rm SNJ}$  localization factor
(second term in Eq.~(\ref{snjtrial})),
which is constructed in such a way that voids 
originated by multiple occupancy of a same site are energetically 
penalized. 
Correlation functions in Eq.~(\ref{snjtrial}) were adopted in the
McMillan, $f(r) = \exp\left[-1/2~(b/r)^{5}\right]$~, and Gaussian, 
$g(r) = \exp\left[-1/2~(a r^{2})\right]$, forms.
The value of the parameters in factors $f$ and $g$  
were optimized variationally at density $\rho = 0.480~\sigma^{-3}$ 
($\sigma = 2.556$~\AA, $b = 1.08~\sigma$ and $a = 10.10~\sigma^{-2}$)  
and kept fixed in the rest of simulations performed at different
densities.  

The technical parameters in our calculations were set in order to 
ensure convergence of the total energy per particle to less than 
$0.02$~K/atom. For instance, the value of the mean population 
of walkers was held to $400$ and the length of the imaginary 
time-step $\Delta \tau$ was $5 \cdot 10^{-4}$~K$^{-1}$~. 
Statistics were accumulated over $10^{5}$ DMC steps
performed after system equilibration, and the approximation used 
for the short-time Green's function 
$e^{-\frac{i}{\hbar} \hat{H} \Delta \tau}$ was accurate
up to order $(\Delta \tau)^3$.~\cite{chin90}

\subsection{Computational strategy}
\label{subsec:strategy}

\begin{figure}
\centerline{
\includegraphics[width=0.8\linewidth]{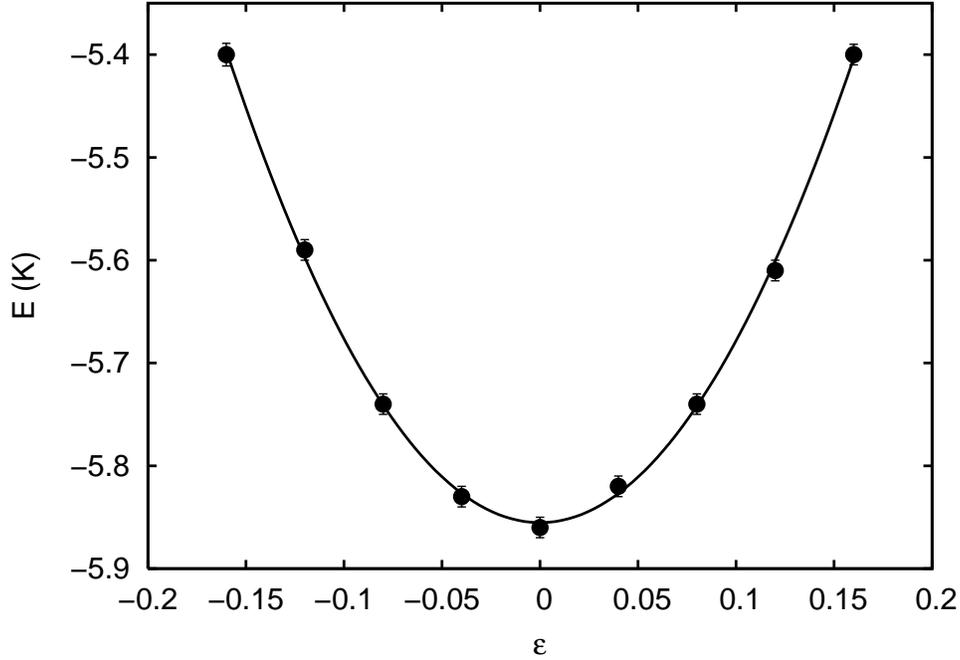}}%
\caption{$C_{44}$ shear energy results obtained in 
         perfect hcp $^{4}$He at density $\rho = 0.480~\sigma^{-3}$.  
         The equilibrium value  
         corresponding to the undistorted hcp structure  
         is quoted at $\epsilon = 0$ and  
         the solid line represents a third-order polynomial 
         fit to the energies.}
\label{fig:fit}
\end{figure}

In order to work out Eq.~(\ref{eq:bulk})   
we used the bulk modulus volume dependence    
reported in Ref.~[\onlinecite{lutsyshyn10}],
where the equation of state of perfect 
hcp $^{4}$He was already calculated 
employing the DMC method and considering variational 
finite-size corrections to 
the total energy.~\cite{cazorla08a} 
The equilibrium value of the $c/a$ ratio
was found to be constant and equal to $1.63(1)$
over all the pressure range $0 \le P \le 110$~bar. 
(This outcome is consistent 
with previous first-principles results obtained 
by other authors.~\cite{freiman09})
Consequently, the left-hand side of Eq.~(\ref{eq:lnca}) 
vanishes and the solution to the fifth-order equation 
system defined by Eqs.~(\ref{eq:bulk})-(\ref{eq:c44})
is
\begin{eqnarray}
C_{11} = K + C_{66} + \frac{1}{18} C_{0} \nonumber \\ 
C_{12} = K - C_{66} + \frac{1}{18} C_{0} \nonumber \\
C_{13} = K - \frac{1}{9} C_{0} \nonumber \\
C_{33} = K + \frac{2}{9} C_{0} \nonumber \\
C_{44} = C_{44}~.
\label{eq:solution}
\end{eqnarray} 

The simulation box used in our pure shear calculations
contains $200$ He atoms and was generated by 
replicating the hcp unit cell $5$ times along the
${\bf a_{1}}$ and ${\bf a_{2}}$ directions and $4$ 
times along the $c$ axis (see Fig.~\ref{fig:supercell}).
In proceeding so, hexagonal symmetry in our supercell
calculations is guaranteed by construction.
Periodic boundary conditions were imposed 
along the three directions defining the 
edges of the non-orthorombic simulation boxes. 
 
The value of the second derivatives involved in 
Eqs.~(\ref{eq:c0})-(\ref{eq:c44}) were computed
following the next strategy. For each volume and 
pure shear considered, first we calculated the 
total energy per particle in a series of supercells 
generated by incrementally distorting the equilibrium 
geometry according to the transformed translational 
lattice vectors~(\ref{eqn:c0vec})-(\ref{eqn:c44vec}). 
Up to eight different and equally spaced shear 
increments (e.g., $\epsilon$, $\gamma$ and $\phi$)
were considered for each volume, taking both positive 
and negative values. 
Subsequently, the series of shear-dependent total 
energies so obtained at fixed volume were fitted to a third-order 
polynomial function of the form $f(x) = a + bx^{2} + cx^{3}$. 
In all the cases, we found that the optimal $a$, $b$ and 
$c$ values reproduced the series of calculated
total energies per particle within their statistical errors 
(see Fig.~\ref{fig:fit}).         
Regarding total energy shifts correcting for the 
finite-size effects, no variational energy corrections 
were considered in the pure shear calculations. 
The reason for this is convenience since finite-size 
energy corrections are not expected to depend on 
the small shear distortions considered in 
this work, and consequently they do not contribute 
to the value of the second derivatives 
involved in the calculation of the elastic constants. 
We shall stress that the volume of the simulation 
cell in pure shear calculations is kept fixed in  
constrast to bulk modulus and 
$\partial \ln{(c/a)} / \partial V$ calculations, in which 
accurate finite-size corrections to the energy are 
certainly required.

\section{Results and Discussions}
\label{sec:results}

\subsection{Elastic constants}
\label{subsec:elasticresults}

\begin{figure}
\centerline
        {\includegraphics[width=0.8\linewidth]{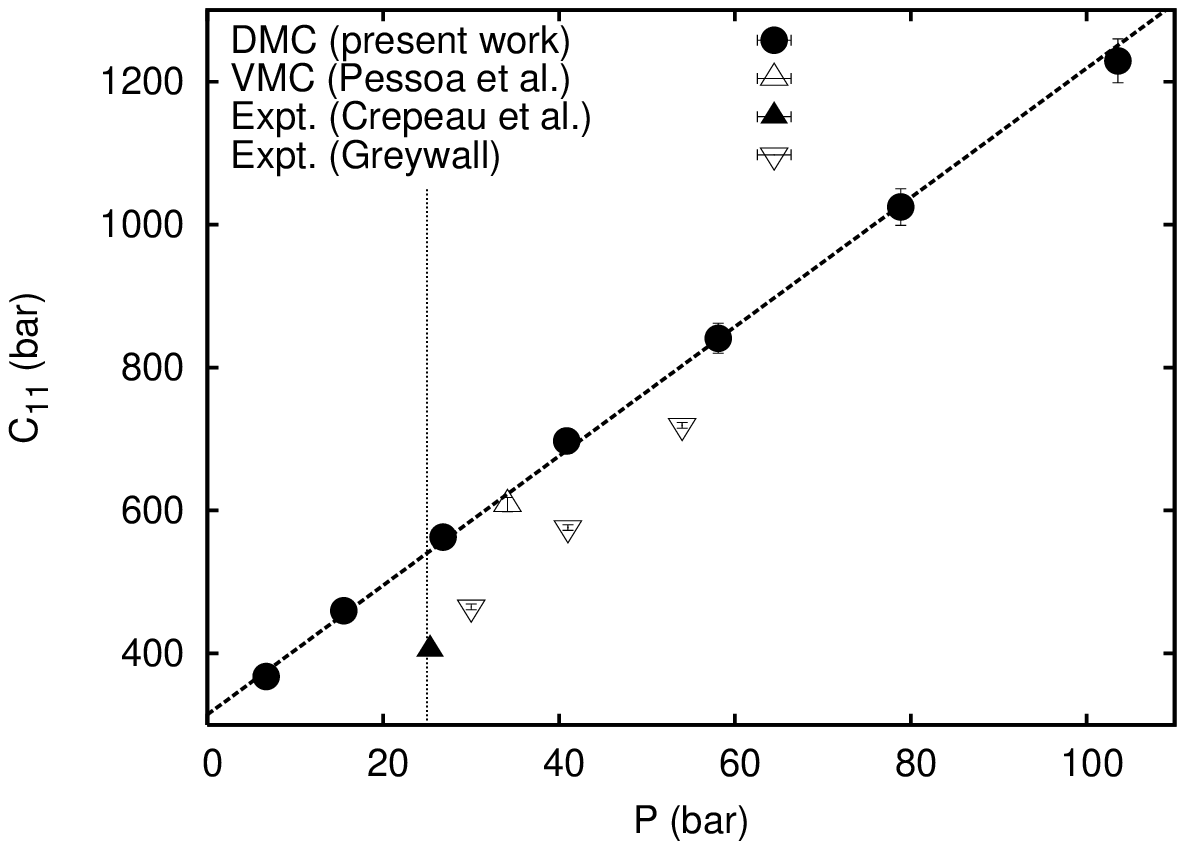}}%
        {\includegraphics[width=0.8\linewidth]{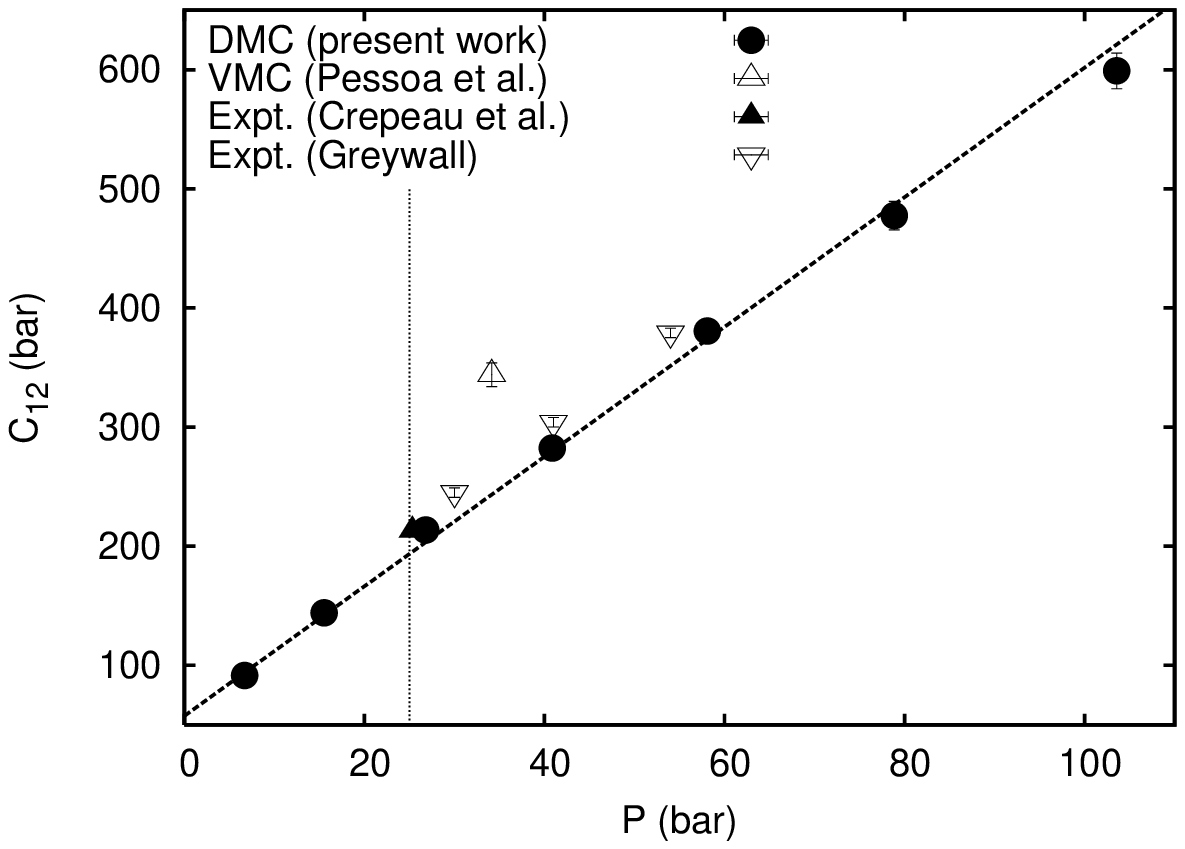}}%
        \caption{Zero-temperature $C_{11}$ and $C_{12}$ elastic constants of
         perfect hcp $^{4}$He as a function of pressure. Previous
         variational Monte Carlo (VMC) calculations~[\onlinecite{pessoa10}] and experimental 
         data~[\onlinecite{crepeau71}]~[\onlinecite{greywall71}] are shown 
         for comparison. The vertical dotted line represents the
         zero-temperature freezing pressure of $^{4}$He and the straight dashed
         lines are linear fits to the DMC results (see text).  
         } 
\label{fig:c11-c12}
\end{figure}

\begin{figure}
\centerline
        {\includegraphics[width=0.8\linewidth]{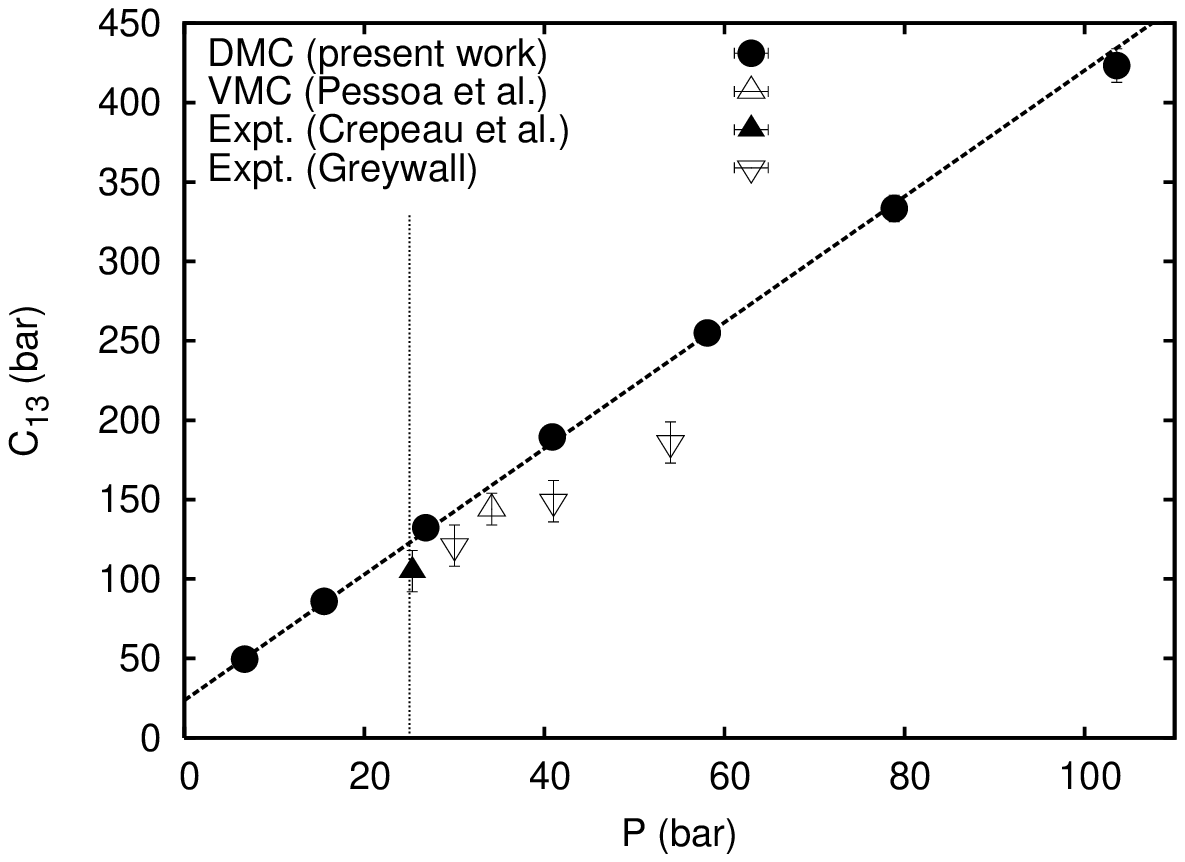}}%
        {\includegraphics[width=0.8\linewidth]{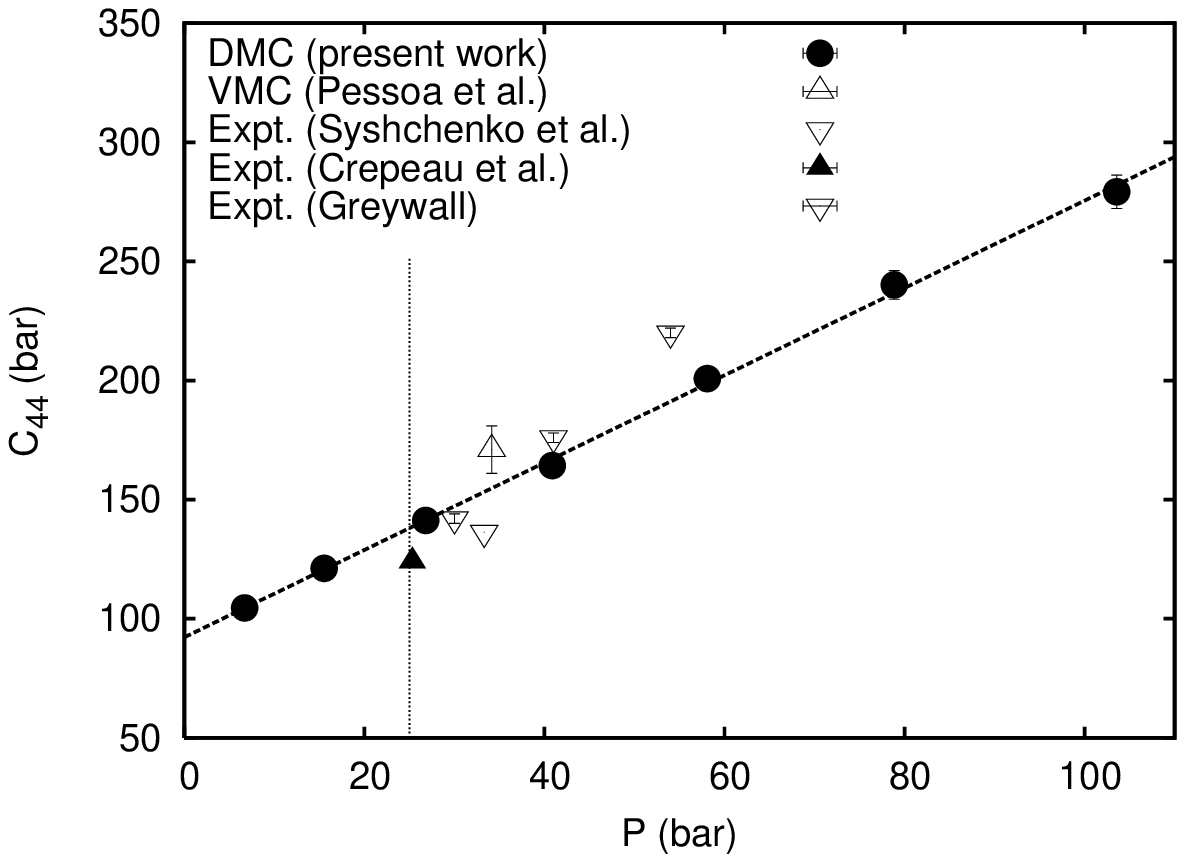}}%
        \caption{Zero-temperature $C_{13}$ and $C_{44}$ elastic constants of
         perfect hcp $^{4}$He as a function of pressure. Previous
         variational Monte Carlo (VMC) calculations~[\onlinecite{pessoa10}] and experimental
         data~[\onlinecite{crepeau71}]~[\onlinecite{greywall71}]~[\onlinecite{syshchenko09}] 
         are shown for comparison. The vertical dotted line represents the
         zero-temperature freezing pressure of $^{4}$He and the straight dashed
         lines are linear fits to the DMC results (see text).
	}
\label{fig:c13-c14}
\end{figure}

\begin{figure}
\centerline
        {\includegraphics[width=0.8\linewidth]{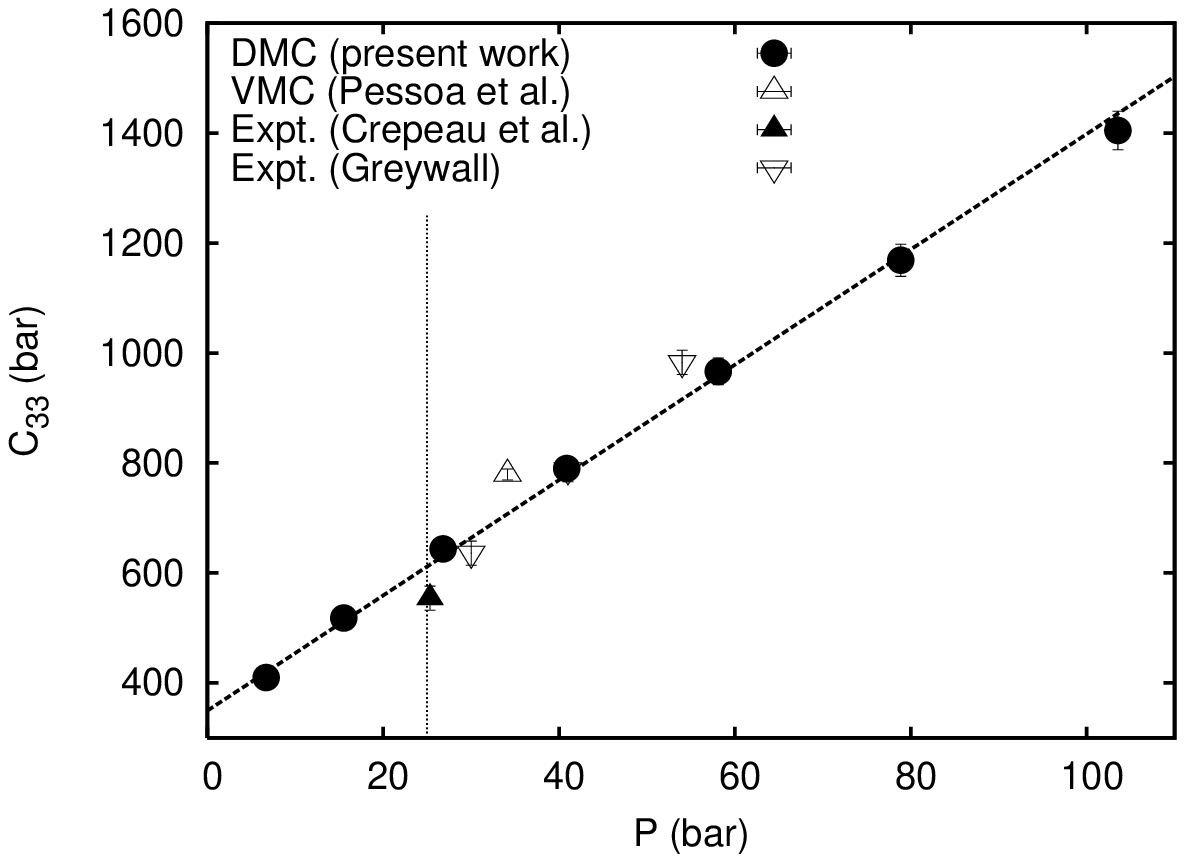}}%
        {\includegraphics[width=0.8\linewidth]{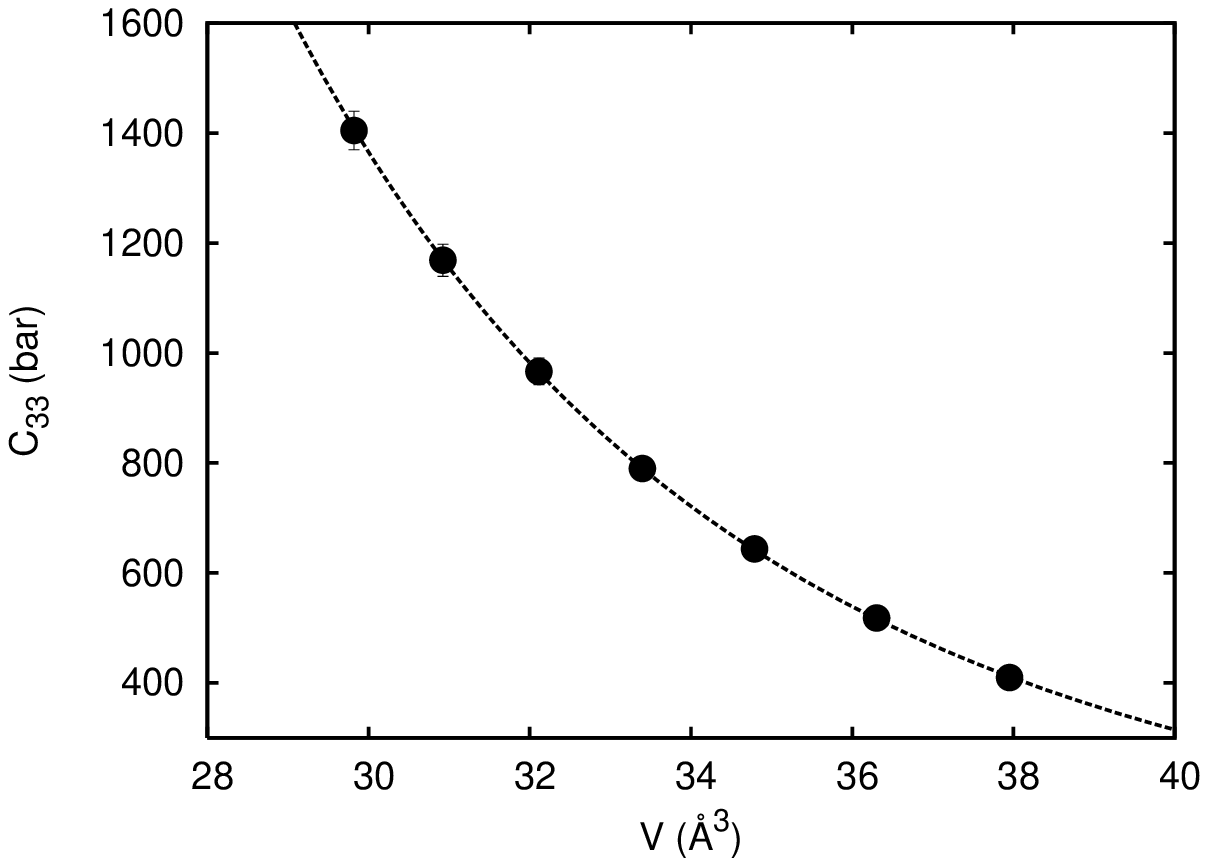}}%
        \caption{ \emph{Top}: Zero-temperature $C_{33}$ elastic constant of
         perfect hcp $^{4}$He as a function of pressure. A previous
         variational Monte Carlo (VMC) calculation~[\onlinecite{pessoa10}] and experimental
         data~[\onlinecite{crepeau71}]~[\onlinecite{greywall71}]
         are shown for comparison. The vertical dotted line represents the
         zero-temperature freezing pressure of $^{4}$He and the straight dashed
         line is a linear fit to the DMC results (see text).  
         \emph{Bottom}: Dependence of the $C_{33}$ elastic constant on volume.
         The dashed line represents a power law fit to the DMC results from which 
         the value of the corresponding Gr\"{u}neisen parameter is obtained (see text).      
         }
\label{fig:c33-gru}
\end{figure}

In Figures~\ref{fig:c11-c12},~\ref{fig:c13-c14} and~\ref{fig:c33-gru}, we 
show the pressure dependence of the five elastic constants of perfect hcp 
$^{4}$He as obtained in our calculations. The error bars $\delta C_{ij}$
in our results, steming from both the statistical uncertainties 
of the energies and corresponding third-order polynomial fits, 
typically amount to $\delta C_{ij} / C_{ij} \sim 2 \%$~.  
We found that the pressure variation of all five elastic constants  
is practically linear within all the studied range. Consequently,
we performed fits of the form 
$C_{ij} (P) = a_{ij} + b_{ij} P$ to our results 
(see Figures~\ref{fig:c11-c12},~\ref{fig:c13-c14} and~\ref{fig:c33-gru}), 
and obtained a series of $a_{ij}$ and $b_{ij}$ coefficients
that we quote in Table~I.  
We note that average variance values obtained in the 
reduced-$\chi^{2}$ tests corresponding to our fits were
always smaller than $2$. 

\begin{table}
\begin{center}
\begin{tabular}{c  c  c  c  c  c}
\hline
\hline
$\qquad \qquad $       & \qquad $C_{11}$ \qquad & \qquad $C_{12}$ \qquad  & \qquad  $C_{13}$ \qquad  &  $\qquad C_{33}$ \qquad  &  \qquad  $C_{44}$ \qquad  \\ 
\hline
\hline
$a_{ij}$  &  $314.26$  &  $57.45 $  &  $23.59 $  &  $349.22$  &  $92.19 $ \\ 
$b_{ij}$  &  $9.05  $  &  $5.44  $  &  $3.97  $  &  $10.49 $  &  $1.83  $ \\ 
\hline
\hline
\end{tabular}
\end{center}
\caption{Value of the parameters obtained in the linear fits to our 
         $C_{ij} (P)$ results (see text). $a_{ij}$'s are expressed in units
         of bar.}
\end{table}

Comparison between our DMC calculations, previous variational
Monte Carlo (VMC) results and experimental data is provided 
also in Figures~\ref{fig:c11-c12}-\ref{fig:c33-gru}~. 
VMC results reported at $P \sim 34$~bar~\cite{pessoa10} 
have been obtained by Pessoa \emph{et al.} using a shadow wave 
function model (SWF).~\cite{vitiello88} This type of trial wave 
function correctly accounts for the atomic Bose-Einstein statistics, 
is translationally invariant and so far it has yielded the most 
accurate variational description of solid helium.~\cite{moroni98} 
Arguably, Pessoa's VMC predictions are in fairly good agreement 
with our DMC results since in general the 
relation $|C_{ij}^{\rm VMC} - C_{ij}^{\rm DMC}|/C_{ij}^{\rm DMC} \le 10 \%$ 
is fulfilled. (This inequality is only violated by 
$C_{12}$ however in that case measurements 
appear to follow closely our results.) 
Recalling that evaluation of $C_{ij}$'s 
requires from the computation of total energy second     
derivatives, it can be said that the satisfactory 
DMC-VMC agreement found further corroborates the excellent 
variational quality of the SWF model. 

Regarding the experimental data taken from 
Refs.~[\onlinecite{crepeau71,greywall71}], we  
also find good agreement with them 
(see Figures~\ref{fig:c11-c12}-\ref{fig:c33-gru}). 
The sound-velocity measurements performed 
by Crepeau and Greywall involved high-quality single 
helium crystals whose basal plane orientations were 
accurately determined using x-rays. Consequently, our 
modest discrepancies with Crepeau and Greywall's data 
are very likely to be originated by thermal effects 
given that the temperature conditions in those 
experiments were $T \sim 1$~K. 
Reassuringly, our $C_{44}$ results reproduce closely 
recent $^{4}$He shear modulus measurements performed by 
Beamish \emph{et al.} at just few mK
(see Figure~\ref{fig:c13-c14}).~\cite{day07,syshchenko09}
We will comment again on this issue in Section~\ref{subsec:soundresults} 
however it can be already claimed that the manifested 
overall good accordance between our $C_{ij}$ calculations 
and $30 \le P \le 60$~bar experiments appears to 
endorse the reliability of our computational 
approach.    

Another quantity of interest in the study of crystal 
elasticity is the Gr\"{u}neisen parameter $\hat{\gamma}$.
Essentially, this parameter quantifies how atomic 
vibrations in a crystal are affected by changes in volume. 
This quantity is customarily defined by 
$\hat{\gamma} = \left( V / C_{v} \right) \beta K$,
where $C_{v}$ stands for the specific heat and $\beta$ for
the coefficient of the thermal expansion. However, this 
definition is not practical for low temperature calculations 
since in general quantities $C_{v}$ and $\beta$ tend to zero 
in a similar trend near $T = 0$, and thus leads to an indetermination. 
Alternatively, Klein \emph{et al.}~\cite{klein70} 
derived a Gr\"{u}neisen parameter expression 
that is valid in the zero-temperature limit and which depends 
on the individual vibrational frequency modes. 
Specifically, Klein's expression can be reformulated in terms 
of the elastic constants as~\cite{greywall71}
\begin{equation}
\gamma_{ij} = - \frac{1}{2} \frac{\partial \ln{C_{ij}}}{\partial \ln{V}} - \frac{1}{6}~.
\label{eq:gruneisen}
\end{equation} 
Naturally, the volume dependence of each $C_{ij}$ 
elastic constant can then be fitted to a function of the 
form
\begin{equation}
C_{ij} (V) = A \left( \frac{V}{V_{0}}\right)^{-\left( \frac{1}{3} + 2\gamma_{ij} \right)}~,
\label{eq:volumefit}
\end{equation}
so that one can readily obtain the value of the 
corresponding $\gamma_{ij}$ parameter.
We proceeded in this way using the $C_{ij}$ results 
obtained in our $0 \le P \le 110$~bar simulations 
(see Figure~\ref{fig:c33-gru}) 
and got $\gamma_{11} = 2.34~(5)$~, $\gamma_{12} = 3.69~(5)$~, 
$\gamma_{13} = 4.23~(5)$~, $\gamma_{33} = 2.70~(5)$ and
$\gamma_{44} = 1.91~(5)$~, where the numerical uncertainties 
are expressed within parentheses. The averaged 
Gr\"{u}neisen parameter  
$\hat{\gamma} = \frac{1}{5} \sum \gamma_{ij}$ corresponding to 
our results is $2.67~(5)$, where 
the summation runs over indexes $11$, $12$, $33$, $44$ and $66$ 
($\gamma_{66} = 2.70$ as obtained from $C_{66}$)
because the respective elastic constants are the quantities which are 
directly measured in sound-velocity experiments.~\cite{greywall71}   
Our $\hat{\gamma}$ value compares very well with Greywall's 
experimental result of $2.7$, however we note that in our calculations
the value of the dispersion $\frac{1}{5} \sum |\gamma_{ij} - \hat{\gamma}|$ 
is non-zero. 
It is worth comparing the value of the zero-temperature Gr\"{u}neisen 
parameter of solid helium to that of other rare-gas species.
We know from Ref.~[\onlinecite{lurie73}] 
that $\hat{\gamma}$ is $2.5$ in Ne, $2.7$ in Ar, 
$2.7$ in Kr and $2.5$ in Xe. Consequently, 
the elastic constants of all five noble gases 
will vary very similarly upon a same volume change. 
The same conclusion, however, does not apply to  
pressure since the bulk modulus of each element  
is appreciably different from that of the others.

In order to quantify the importance of quantum effects
in our study, we computed the contribution of 
the potential and kinetic energies to the shear modulus
($C_{44}^{p}$ and $C_{44}^{k}$, respectively). 
For this, we carried out simulations at density 
$\rho = 0.480~\sigma^{-2}$ in which the \emph{exact} value 
of the second derivative of the potential energy $E^{p}$ with 
respect to strain was calculated using 
the pure estimator technique.~\cite{casulleras95}     
The kinetic energy contribution to the shear 
modulus $C_{44}^{k}$ was subsequently obtained by subtracting
the quantity   
$C_{44}^{p} = \frac{1}{V_{0}}\left( \frac{\partial^{2} E^{p}}{\partial \phi^{2}} \right)_{V=V_{0}}$ 
to $C_{44}$.(We checked that the strain dependence of 
$E^{k}$ could also be accurately fitted 
to a third-order polynomial function). 
In fact, the $T = 0$ value of $C_{44}^{k}$ in a classical 
crystal exactly amounts to zero since the atoms 
there remain frozen in their perfect lattice positions  
(that is, $C_{44}^{p} = C_{44}$).
Even in the case of considering quasi-harmonic 
zero-point motion corrections to $C_{44}$, $C_{44}^{k}$ is 
not expected to depart significantly from zero.       
In contrast, we found that in perfect hcp $^{4}$He   
$C_{44}^{p} / C_{44}$ amounts to $68~\%$, or conversely,  
$C_{44}^{k} / C_{44} = 32~\%$. This last result 
quantifies the quantum nature of solid helium's elasticity 
and demonstrates the inability of classical and quasi-harmonic
approaches for reproducing it.

\subsection{Sound velocities}
\label{subsec:soundresults}

\begin{figure}
\centerline{
\includegraphics[width=0.8\linewidth]{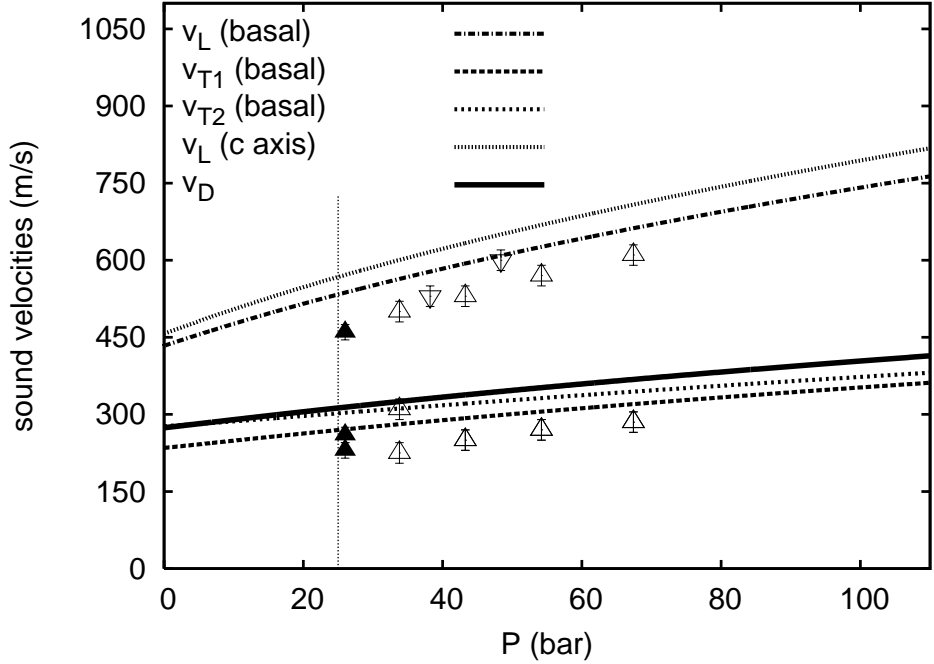}}%
\caption{Pressure dependence of the longitudinal~(L) and 
transvere~(T) sound velocities of hcp $^{4}$He along 
its corresponding $c$-axis and basal plane.
$v_{D}$ represents the averaged Debye velocity (see text).
Basal sound-velocity data from Refs.~[\onlinecite{greywall71}]~($\triangle$),
[\onlinecite{crepeau71}]~($\blacktriangle$) and~[\onlinecite{wanner70}]~($\triangledown$) 
are shown for comparison. The vertical dotted line represents the
zero-temperature freezing pressure of $^{4}$He.
}
\label{fig:sound}
\end{figure}

\begin{figure}
\centerline{
\includegraphics[width=0.8\linewidth]{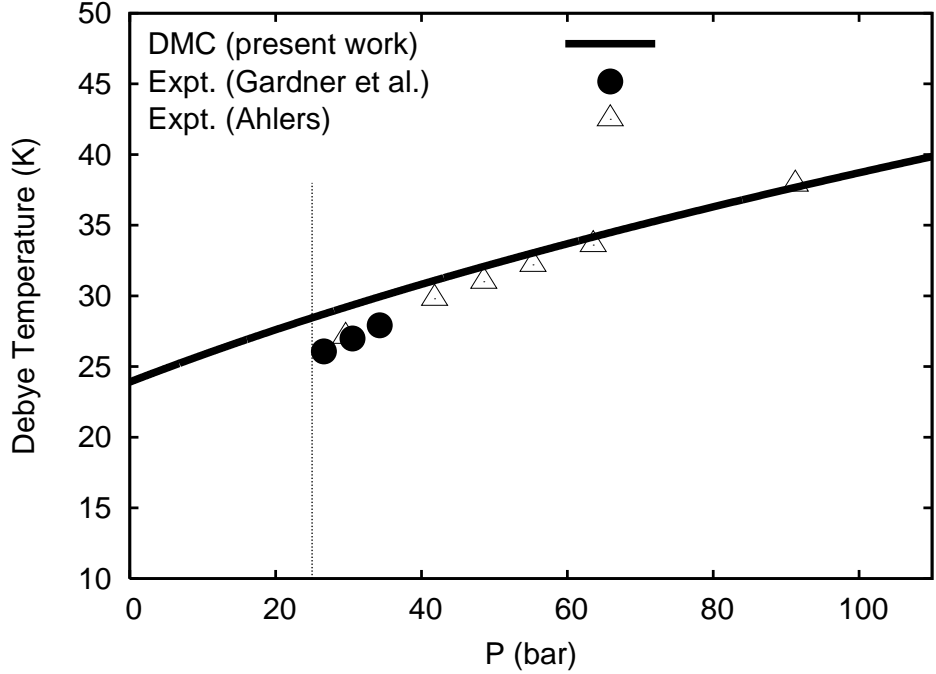}}%
\caption{$T = 0$ Debye temperature of hcp $^{4}$He
as a function of pressure (solid line). Experimental 
results from Refs.~[\onlinecite{gardner73}] and~[\onlinecite{ahlers70}] 
are shown for comparison. The vertical dotted line represents the
zero-temperature freezing pressure of $^{4}$He, and the thickness
of the line corresponds to the uncertainty associated to our calculations 
(e.g. $\delta \Theta_{D} / \Theta_{D} \sim 1 \%$)~.} 
\label{fig:debye}
\end{figure}

Sound velocities in solids, either longitudinal or tranverse, 
depend on their direction of propagation. 
In crystals with hexagonal symmetry two main propagation modes are identified, 
one along the $c$-axis (defined by vector ${\bf a_{3}}$ in Section~\ref{subsec:elasticcte}) 
and the other contained within the basal plane (defined by vectors ${\bf a_{1}}$ and ${\bf a_{2}}$ 
in Section~\ref{subsec:elasticcte}). The relationships between the elastic constants
and sound velocities in hcp crystals are~\cite{musgrave54,goldman79} 
\begin{eqnarray}
v_{L} = \left( C_{33} / \rho \right)^{1/2}~\nonumber \\
v_{T1} = \left( C_{44} / \rho \right)^{1/2}~\nonumber \\
v_{T2} = \left( C_{44} / \rho \right)^{1/2}
\label{eqn:cvel}
\end{eqnarray}
along the $c$-axis, and
\begin{eqnarray}
v_{L} = \left( C_{11} / \rho \right)^{1/2}~\nonumber   \\ 
v_{T1} = \left( C_{66} / \rho \right)^{1/2}~\nonumber  \\
v_{T2} = \left( C_{44} / \rho \right)^{1/2}  
\label{eqn:basalvel}
\end{eqnarray}
within the basal plane.
     
In Fig.~\ref{fig:sound}, we plot the pressure dependence of 
the tranverse and longitudinal sound velocities of hcp $^{4}$He 
as obtained from our $C_{ij}$ results reported in 
Section~\ref{subsec:elasticresults}. The error bars  
in our results, not shown in the figure, are 
$\delta v_{L,T} / v_{L,T} \sim 1 \%$~.  
It is observed that at compressions far beyond freezing all sound velocities 
increase almost linearly with pressure. In contrast, the longitudinal $c$-axis 
and basal components appear to follow a certain power-law within the 
low-density interval $0 \le P \le 25$~bars. Certainly, the nature of 
the sound propagation modes in metastable solid $^{4}$He, either at 
positive or negative pressures, is poorly understood at present 
in spite of its fundamental physical interest.~\cite{maris09} 
It is our aim to report with detail on this topic in elsewhere 
so we leave discussions on this matter out of this work.  

Experimental longitudinal and tranverse basal sound velocities are 
shown for comparison in Fig.~\ref{fig:sound}. The agreement 
between those measurements and our predictions is generally good  
(in fact, as good as claimed in the previous section for 
the elastic constants). 
Specifically, our predicted sound velocities 
are systematically a bit larger than those 
values reported by Wanner~\cite{wanner70}, Crepeau~\cite{crepeau71} and
Greywall.~\cite{greywall71} Such a systematic 
overestimation is consistent with our previous suggestion 
that certain thermal effects, non-reproducible by our
simulations, could be affecting the experiments. 
As a fact of matter, the less rigid a material becomes by effect 
of temperature, the more slowly sound waves propagate across 
of it.
 
In order to provide a more meaningful comparison between our 
zero-temperature results and experiments, we computed the $T = 0$ Debye 
temperature of $^{4}$He $\Theta_{D}$. The zero-temperature $\Theta_{D}$ 
of a crystal can be easily extrapolated from lattice heat-capacity 
measurements performed at low temperatures, and results for this quantity 
have already been reported for helium over a wide pressure 
range.~\cite{gardner73,ahlers70}
The definition of the $T = 0$ Debye temperature is   
\begin{equation}
\Theta_{D} = \frac{2\pi \hbar}{k_{B}} \left( \frac{3}{4 \pi V}  \right)^{\frac{1}{3}} v_{D}~,
\label{eq:debye}
\end{equation}
where $V$ is the volume per atom and $v_{D}$ the Debye velocity.  
This velocity is given by 
\begin{equation}
\frac{1}{v_{D}^{3}} = \frac{1}{3} \left( \frac{1}{\overline{v}_{L}^{3}} + 2\frac{1}{\overline{v}_{T}^{3}}\right)~,
\label{eq:debyevel}
\end{equation} 
where the average velocities $\overline{v}_{L}$ and $\overline{v}_{T}$ are 
defined by
\begin{equation}
\frac{1}{\overline{v}_{L,T}^{3}} = \langle \frac{1}{v_{L,T}^{3}} \rangle~,
\label{eq:averagevelocities}
\end{equation}
and the $\langle \cdots \rangle$ brackets denote angular averages of 
the longitudinal and tranverse velocities. In our case, we have
approximated the angular averages in Eq.~\ref{eq:averagevelocities} by  
\begin{equation}
\frac{1}{\overline{v}_{L}^{3}} \approx \frac{1}{2} \left( \frac{1}{v_{L,b}^{3}} \right) + \frac{1}{2} \left( \frac{1}{v_{L,c}^{3}} \right)
\label{eq:approx1}
\end{equation}
and
\begin{equation}
\frac{1}{\overline{v}_{T}^{3}} \approx \frac{1}{3} \left( \frac{1}{v_{T1,b}^{3}} \right) + \frac{1}{3} \left( \frac{1}{v_{T2,b}^{3}} \right) + \frac{1}{3} \left( \frac{1}{v_{T,c}^{3}} \right)~,
\label{eq:approx2}
\end{equation}
where index $b$ stands for basal and index $c$ for $c$-axis.~\cite{aclaration2}

In Fig.~\ref{fig:debye}, we plot our results for the 
zero-temperature $\Theta_{D}$ of hcp $^{4}$He and experimental data taken 
from Refs.~[\onlinecite{gardner73,ahlers70}]. In fact, excellent agreement 
between Gardner and Ahlers measurements and our calculations is observed. 
This last result appears to further ratify our previous suggestion 
that, once thermal effects are corrected for, our $T = 0$ 
elastic constants and deriving quantities predictions closely reproduce 
experiments.  
Finally, we note that the pressure variation of $\Theta_{D}$ is very
similar to that observed in the longitudinal sound-velocity components
of helium, namely almost linear at high compressions and of power-law type  
at low densities.

\section{Summary and Perspectives}
\label{sec:summary}

We have developed a fully quantum computational strategy to accurately 
calculate the zero-temperature elastic constants of perfect hcp 
$^{4}$He under pressure. 
Our diffusion Monte Carlo results are shown to be consistent 
with low-$T$ sound-velocity measurements and previous variational
first-principles calculations. It is found that  
all five non-zero elastic constants of helium vary linearly
with pressure within the range $0 \le P \le 110$~bar, 
and we have provided an accurate parametrization of each
of them. The Gr\"{u}neisen parameters, sound velocities and 
$T = 0$ Debye temperature of solid helium have been also determined 
and compared to experimental data. 
The computational method introduced in this work is totally 
general so that it can be used for the study 
of any other hcp quantum solid appart from helium (e.g. H$_{2}$), 
and/or be conveniently altered in order to investigate other 
crystal structures (e.g. face centered and body centered cubic).

It is our intention to analyze the elastic behavior of $^{4}$He at 
negative pressures using the computational technique described here. 
In doing this, we expect to be able to determine 
its spinodal density limit (that is, the density at which the 
elastic constants vanish) rigorously, and also
characterize the pressure dependence of the tranverse and basal sound 
velocities near it. Also we are interested in applying our formalism 
to the study of the ground-state of defective hcp $^{4}$He (for instance, by 
introducing vacancies), where the supersolid state of matter clearly 
manifests. In doing this, we expect to gather quantitative knowledge on 
the relationship (if any) between elasticity and supersolidity and 
so to help to understand the origins of recent shear modulus observations. 
Work in these directions is already in progress.

\begin{acknowledgments}
The authors acknowledge partial financial support from the  
DGI (Spain) Grant No.~FIS2008-04403 and Generalitat de Catalunya 
Grant No.~2009SGR-1003.
\end{acknowledgments}


\begin{thebibliography}{30}
\bibitem{aclaration} By related quantities is meant the expected value of 
                     operators $\hat{A}$ that commute with the Hamiltonian, namely
                     $[\hat{A}, \hat{H}] = 0$~. It is also possible to 
                     obtain virtually exact results for some 
                     particular $[\hat{A}, \hat{H}] \neq 0$ operators
		     by using forward walking based techniques (see Refs.~[\onlinecite{barnett91}]
                     and [\onlinecite{casulleras95}]).
\bibitem{aclaration2} In view of the experimental results reported in 
                      Fig.~10 of Ref.~[\onlinecite{greywall71}], this 
                      approximation can be considered as certainly good.
\end{thebibliography}
\end{document}